# Activated Layered Magnetism from Bulk TiN

Chiung-Yuan Lin,[1]* Szu-Wen Yang,[1] Keng-Liang Ou,[2,3,4,5] Barbara A. Jones[6]*

**Affiliations**

[1]Department of Electronics Engineering and Institute of Electronics, National Chiao Tung University, Hsinchu, Taiwan.

[2]Department of Dentistry, Taipei Medical University Hospital, Taipei 110, Taiwan

[3]Department of Dentistry, Taipei Medical University-Shuang Ho Hospital, New Taipei City 235, Taiwan

[4]3D Global Biotech Inc., New Taipei City 221, Taiwan

[5]Department of Dentistry, Cathay General Hospital, Taipei 106, Taiwan

[6]IBM Research-Almaden, San Jose, California 95120-6099, USA.

*Corresponding author. Email: genelin@mail.nctu.edu.tw (C.-Y. L.); bajones@us.ibm.com (B.A. J.)

**Abstract**

The novel properties of a uniaxially-expanded TiN bulk arising from increasing the layer spacing from equilibrium are explored using a first-principles approach. We reveal a novel nonmagnetic-magnetic transition from a TiN bulk to its monolayer. We also investigate the electronic and magnetic structures of a few TiN atomic layers. We find that the bilayer and trilayer, like the TiN bulk, are nonmagnetic poor metals. On the other hand, the monolayer TiN is found to carry a magnetic moment on its Ti atoms, and likely be a semiconductor. The unpaired electron giving rise to magnetism on Ti is primarily in the orbital perpendicular to the layers, and we find it is freed to give rise to magnetism when the layers are slightly separated. We find two different antiferromagnetic states possible on the monolayer, as well as one ferromagnetic, with one of the antiferromagnetic being the lowest energy. The exchange couplings between Ti atoms in such a monolayer are calculated to be antiferromagnetic for both the nearest-neighbor and next-nearest-neighbor sites. We also analyze the binding nature of both the monolayer and bilayer TiN by searching for the predominant binding orbitals.

**MAIN TEXT**

**Introduction**

Transition-metal nitrides have extraordinary material properties: high melting temperature, high stiffness, high Curie temperature, which provide a variety of surface coating applications. They are mostly conductors in their electrical properties, and some of their thin films even exhibit superconductivity. As a result they are extensively studied by scientists. Among them titanium nitride (TiN) draws especially great attention. Its high hardness and corrosion-resistance make it one of the best wear-resistant coatings of cutting and threading tools. It can also serve as conductive barrier layers in semiconductor devices for its good conductivity and high diffusion barrier. Other applications of TiN include coating costume jewelry with its metallic gold color and transparent thin films of selective wavelengths. Besides the variety of coating applications, its microscopic magnetism is also of potential considerable interest. Bulk TiN has no magnetism, not even microscopic antiferromagnetism (*1*). On the other hand, a Ti atom carries a magnetic moment as a free atom and as an adatom binding to N or O on the surface (*2-4*). This insight leads one to expect that a few atomic layers of TiN may exhibit magnetism. If one can find such an atomic-scale magnetic ultrathin film, its novel physical properties (magnetic and optical) would have a wide range of applications, such as magneto-optical isolators, sensors, circulators, and information storage. Such atomic-scale structures can also potentially serve as biomedical magnetic coatings, e.g., on the acupuncture needle tip to increase its magnetic stimulation in the acupuncture points. The magnetic-nonmagnetic transition of TiN from a monolayer to the bulk implies that the magnetism of TiN may be tuned by the spacing between its atomic layers. Implementation of such nanostructures could eventually lead to highly sensitive magnetic sensors and actuators (*5,6*).

     It is well known that materials of a few atomic layers can turn into novel electronic structures. Graphene is substantially different from graphite: the stiffest nanomaterial with high thermal conductivity, astonishing carrier mobility, and mean free paths of several micrometers at room temperature (*7,8*). Bulk transition-metal sulfides and selenides $MoS_2$, $WS_2$, $MoSe_2$, and $WSe_2$ are indirect-bandgap semiconductors (*9,10*), while their single atomic layers change to direct bandgaps (*11-13*). Such size effects also widely appear in conventional semiconductors and metals. The band gaps of semiconductor thin films increase due to quantum confinement as the thickness shrinks (*14,15*). Metallic ultrathin films are also affected by quantum confinement when the film thickness is comparable to the Fermi wavelength, leading to unexpected growth behaviors from the interplay between level quantization, charge spilling effects, Friedel oscillations, etc (*16*). All the above illustrate atomic ultrathin layers may possess substantially different electronic structures and transport properties.

     Like a nonmagnetic surface with a Ti adatom on it becomes magnetic at the Ti sites, a nonmagnetic bulk can be expected to turn magnetic by adding specific dopants with a free-atom magnetic moment (*17-19*). Likewise, a 2D material gains magnetic moments that are induced by point defects (*20-23*). Moreover, a nanoribbon can have line-edge magnetism. For example, $MoS_2$ and $WS_2$ can have magnetic moments at the edges of their nanoribbons (*24,25*). The magnetic properties of $WS_2$ nanoribbon can vary for different ribbon widths. Those monolayers are nonmagnetic semiconductors even though composed of magnetic atoms Mo and W, but subsequently their nanoribbons can turn into magnetic metals. In addition to the line edges of two dimensional ribbons, a recent study demonstrates layer-dependent magnetic ordering from a bulk material $CrI_3$ down to its single atomic layer (*26*). Likewise, even though for the material of interest of this study, TiN, no magnetism is found either for its pure bulk or when being implanted with foreign atoms (*27,28*), we expect, based on analogy to the effects of dimensionality in the nanoribbon examples earlier in this paragraph, that its few atomic layers (with its surfaces considered as planes of "vacancies" and/or "defects") are very likely to be magnetic. There is one experimental work studying the mechanical properties of TiN monolayer on steel, with no investigation of its magnetism (*29*). On the other hand, past computational studies of a few atomic layers of TiN include TiN on a MgO substrate (*30*) and free-standing buckled monolayer of TiN (*31*). The former studies the ionically bounded TiN/MgO interface, such as work function, interface energy and adhesion energy, and concludes that the structure benefits development of gate-stack materials in metal-oxide-semiconductor devices. They do not calculate magnetic properties, however. The latter reference shows the buckled monolayer of TiN exhibits no magnetism and possesses unique mechanical properties of auxeticity. Furthermore, it suggests that the halogenations of buckled TiN monolayers are potential photocatalysts for water splitting applications (hydrogen production).

We are thus motivated to study nanolayers of TiN, to see under which conditions magnetism arises in this material. We calculate using density functional theory a TiN monolayer that is formed within TiN bulk by inserting Ar atoms into the interstitials above and below. To further understand the fundamental properties we calculate one to several layers of TiN, as a simplified model system. We calculate the coupling between the Ti atoms and find whether they couple ferro- or antiferromagnetically as well as their coupling strength. We are also interested in the number of atomic layers at which the magnetic-nonmagnetic transition occurs. Moreover, a nonmagnetic bulk and a possibly magnetic single layer for TiN imply that we may expand the lattice only along one axis to lift apart the bulk layers, and expect another magnetic-nonmagnetic transition as the bulk is expanded. Experimentally we would envision these few-layer systems as semi-free-standing unbuckled layers with, for example, Van der Waals coupling to the substrate.

In this work, we compare the TiN of one to a few atomic layers with its bulk by performing first-principles calculations, and explore the transition from one monolayer of TiN to bulk properties. We start by calculating the electronic and magnetic structures of one, two, and three atomic layers of TiN. The magnetic properties include the magnetic moments, ferro- and antiferro- magnetism, the magnetic couplings between Ti atoms, and the spin-resolved partial density of states (PDOS). The spin-resolved PDOS is further analyzed to determine from which particular orbital the magnetism originates. We then analyze the binding nature between atoms to obtain deeper insights into the electronic properties of these layered structures. In order to further look into the nonmagnetic-magnetic transition due to thickness, we expand our study of the magnetism vs. the number of atomic layers to the magnetism of a uniaxially-expanded TiN bulk. Finally, we present our concluding discussion.

## Results

### Comparison of Electronic and Magnetic Properties

We compare the electronic and magnetic properties of the TiN bulk, monolayer, bilayer, and trilayer. Using the constraint-PBE method (*32*), the *U* values of the above three structures are calculated to be 4.0, 7.7, 7.2, and 7.2 eV, respectively. By employing these *U* values, we calculate magnetic moments and densities of states, as shown in Table 1 and Fig. 1. The bilayer and trilayer TiN contain no magnetic moment, like the bulk, and therefore we expect that TiN(001) thin films are nonmagnetic for two and more atomic layers. However, interestingly, in a surface-induced magnetism effect, the monolayer always carries a magnetic moment of 1.11 on its Ti atoms in different magnetic phases (phases to be detailed later), and we conclude the Ti has $S=1/2$ on a TiN monolayer. We further confirm this effect in structures of a surface monolayer separated from bulk by a layer of Ar, as well as a monolayer of TiN sandwiched in bulk by Ar above and below (see the Supplementary Material "Strained monolayer and bilayer"). In all three cases, we find a separated monolayer displays an induced magnetism.

By comparing Ti 3*d* partial densities of states (PDOS), band structures, and magnetic moments with and without spin-orbit coupling, we find the spin-orbit coupling to be negligible on the above-calculated electronic structures and magnetism. We first compare both the density of states (DOS) and the Ti 3*d* partial density of states (PDOS) of the nonmagnetic bilayer and trilayer in Fig. 1A and 1B as referenced from the bulk, and find that the bilayer and trilayer systems not only both exhibit nonmagnetism but also have very similar density of states in their electronic structures. The electrical property of a TiN bulk is experimentally well-known to be a poor metal (*33*). To qualitatively determine the electrical properties of the bilayer and trilayer, we combine the Boltzmann transport theory and the DFT-calculated electronic structures to find the conductivity divided by the relaxation time $\sigma/\tau$, whose square root is formally the thermodynamic root-mean-square of the fermi velocity times the fundamental charge (see details in ref: *34*). We obtain $\sigma/\tau$ as $1.2675924\times10^{21}$, $1.5005080\times10^{20}$, and $2.1061531\times10^{20}$ $\Omega^{-1}\cdot m^{-1}\cdot s^{-1}$ for the bulk, bilayer, and trilayer, respectively. Given the fact that surface scattering effects in the bilayer and trilayer will result in relaxation times longer than the bulk, consequently they have smaller conductivity. We therefore expect that the TiN(001) thin films beyond two atomic layers are also poor metals.

We can explain the origin of the magnetism of the monolayer in the following way. The monolayer has an abrupt unpaired spin-up DOS peak near 1~2eV below the fermi level as shown in Fig. 1C, contributing a magnetic moment of almost one unpaired electron. The Ti PDOS of all five 3*d* orbitals are plotted in Fig. 1D, further showing that the unpaired electron basically occupies the $3d_{z^2}$ orbital. We will interpret this result in more detail in the next subsection. We calculated the electronic structure of the monolayer using PBE+*U*. However, based on the established methodology to obtain the best estimation of the band gap for materials such as $TiO_2$, which involves not only addition of a $U_d$ to the *d*-orbitals of the transition metal but also another $U_p$ to the *p*-orbitals of the first-row reactive non-metal, we add a *U* term to the p-orbital of N in the monolayer. We see that the bandgap decreases from 1.25eV to 1.11eV as $U_p$ is increased from 0 to 6eV. Based on this analysis, we expect the bandgap of TiN monolayers lies between 1.11 and 1.25eV. This bandgap would make TiN monolayers to be semiconducting.

The magnetic monolayer exhibits rich magnetic phases by different alignments of its atomic spins. There is a ferromagnetic (FM) phase. Two antiferromagnetic (AFM) phases are also found, AFM1 and AFM2. The former consists of opposite atomic spins between nearest-neighbor Ti sites (Fig. 2A), while the latter has alternating spin orientations of (110) atomic-spin stripes (Fig. 2B). We also explore the strength of the spin couplings as follows. The exchange couplings between nearest-neighbor and next-nearest-neighbor Ti atoms have been extracted to be $J_1$=8.68 meV and $J_2$=2.54 meV, both being antiferromagnetic. Notice that the nearest-neighbor coupling is roughly four times that of the next-nearest neighbor one, as expected. The energy differences between different magnetic phases provide an estimate of temperatures of magnetic phase transitions. In fact, the monolayer has its AFM1 phase 3.6 meV lower in energy than AFM2 (equivalent to a temperature ~ 40K) per unit cell, and these two AFM phases are 17.4meV (equivalent to a temperature of ~ 200K) and 13.8meV lower than the FM phase, respectively. Although the two antiferromagnetic phases are not too far apart in energy, clearly the AFM1 phase is the ground state, especially compared to the FM state.

In summary, we show that the TiN(001) monolayer is an antiferromagnetic semiconductor, the unpaired electron residing primarily in the Ti $3d_{z^2}$ orbital. However, in a sudden transition, any other numbers of layers become nonmagnetic poor conductors. (In the Supplementary Materials are further comparisons of magnetism with Ti and N binding on material surfaces.) In the rest of this paper we explore in detail the magnetic phases of the monolayer, how the magnetic/nonmagnetic transition arises, and, how it can be controlled.

**Binding Orbitals**

In order to understand the binding natures of the mono- and bilayer TiN, we search for the particular orbitals that dominate the bonds, and determine their orbital symmetries by both plotting and quantitative decomposition. We find that in the single-layer TiN in its AFM1 phase, the ground state, the peaks of the N $p_x + p_y$ and Ti $d_{x^2-y^2}$ spin-up PDOS align at many energies, as shown in Fig. 3A. Examining in detail the eigenstates at these energies, we focus on one state which exhibits almost pure coupling in the plane, at $E=-2.48$eV. We plot the Kohn-Sham orbital at $E=-2.48$eV in Fig. 3A. This plot shows the strong hybridization of the N $p_x$ and Ti $d_{x^2-y^2}$ atomic orbitals, which lead to binding in the plane.

Next, we show the binding nature out of plane for AFM1. Here the appropriate state we look at corresponds to $E=-1.73$eV, where there are two degenerate orbitals as shown in Fig. 3B. In this out-of-plane case, one state has N $p_z$ and Ti $d_{xz} + d_{yz}$ with N dominant. The other state has Ti $d_{z^2}$ with Ti dominant, with no N present.

In the next row of this figure, we examine the ferromagnetic phase to look at its binding, because it will be instructive when we come to the bilayer. In the ferromagnetic phase, the state of greatest binding is between N $p_x + p_y$ and Ti $d_{x^2-y^2}$ modes (Fig.3C). Notice that the bonds in plane differ from those in AFM1. In Fig. 3D, we plot the orbital of monolayer FM out of plane. Notice the strong similarity to the orbitals out of plane for AFM1 (Fig. 3B). The difference in the number of orbitals in the two figures is solely due to difference of AFM and FM magnetic configurations.

Now we come to the bilayer, where the magnetism approaches zero. Here the interlayer binding nature is the most interesting feature to look for. There is a broad co-peak of the N $p_z$ and Ti $d_{z^2}$ PDOS ranging from -4.8 to -3.3eV, as shown in Fig. 3F. The Kohn-Sham orbital at $E=-4.45$eV is plotted in Fig. 3F, and shows that the interlayer binding is N $p_z$ plus Ti $d_{z^2}$ and is very strong. We also plot the corresponding quantities for the in-plane binding for a bilayer, and this is shown in Fig. 3E. Notice the interesting correspondence of these images to those of the monolayer ferromagnetic along-the-plane case. In contrast, the out-of-plane orbitals for the bilayer look very different from the monolayer case, because of the strong hybridization between layers in the bilayer case. The binding for thicker layers will look similar to those of the bilayer.

**Magnetic Transition from the nonmagnetic bulk to the magnetic monolayer**

One way to consider controlling the magnetic transition between the TiN bulk and monolayer is to slowly lift apart the bulk layers originally stacked at the equilibrium lattice spacing until each bulk layer becomes far apart enough to reasonably mimic one monolayer. The details of the magnetic moments at the intermediate layer spacings bring us closer to such a magnetic-nonmagnetic transition.

In principle, one should use a varying $U$ to the above lattice-expansion calculations, with the $U$ value calculated by the constraint-PBE method on the fly. However, besides being computationally expensive, brute-force calculations of the $U$ values with varying layer spacing often encounter convergence problems at the intermediate layer spacings of rapidly varying magnetic moments. As an alternative, we calculate the uniaxially-expanded bulk at three fixed *U* values: the bulk (4eV), the monolayer (7.7eV), and their average (5.8eV).

We calculate the expanded TiN bulk in the atomic-spin configuration in Fig. 4A. (AFM of alternating layers). The calculated magnetic moments are plotted in Fig. 4B, as functions of the layer spacing along the expanded direction, with all three *U* values, in each plot as shown in the legends. In Fig. 4C we zoom in on the transition region to show more detail. In the cases of all three *U* values we show that the TiN bulk basically remains nonmagnetic with a uniaxial layer spacing below 2.4Å, and undergoes a nonmagnetic-magnetic transition when the spacing is stretched between 2.4 and 2.7Å. As the layer spacing keeps being stretched beyond 3.2Å, the magnetic moments approach the value of a monolayer. Hence, we have here captured the essence of the nonmagnetic-magnetic transition.

**Discussion**

By calculating the electronic structure of ultrathin TiN atomic layers, we reveal the exotic electronic and magnetic transitions by varying the number of layers: A single-layer TiN exhibits physical properties totally different from its multilayers. The latter are nonmagnetic, poor conductors, similar to the bulk, while the former is a magnetic semiconductor with two nearly degenerate AFM phases. The TiN monolayer has one $3d_{z^2}$ unpaired electron per Ti atom, and it may undergo an AFM-FM transition around the temperature of dry ice. Moreover, our further investigations into a bilayer indicate that the above $3d_{z^2}$ magnetic orbital becomes hybridized with the N $p_z$ of the other layer, and consequently its orbital magnetism gets quenched for two layers and more.

The unexpected surface-induced magnetism of a TiN monolayer, contrary to its nonmagnetic bulk, inspires us to further look into the details. We did this in two ways. Our calculations show that inserting Ar into a TiN bulk or on the surface to create a TiN surface layer will produce magnetism on the TiN monolayer. In a second way, we examine a uniaxially-expanded TiN bulk. Such a bulk undergoes, as observed from first-principles calculations, a nonmagnetic-magnetic transition by increasing the layer spacing only along the *c*-axis. This transition is found to have an onset at a 13% expansion, and saturates into monolayer magnetism at 27%.

In fact, our calculations suggest that experimentalists can practically fabricate a magnetic TiN monolayer within a TiN bulk or on the surface by alternatingly depositing molecules and TiN on a substrate or in general by separating layers molecularly. Revealing the unusual magnetic transition induced by uniaxial expansion in a computational study provides innovative directions to further engineer the magnetism of TiN and develop applications of great impact in biomedical coating of tunable magnetism, magnetoresistive sensors, and magnetostrictive ultrathin-film actuators.

**Materials and Methods**

TiN bulk crystallizes in the rocksalt structure, with a lattice constant we calculate to be 4.26Å, in good agreement with the experimental value 4.24Å. We then construct its atomic monolayer, bilayer and trilayer. These TiN atomic ultrathin layers are all chosen in the (001) direction for simplicity, modeled as periodic slabs separated by vacuum of at least 20Å. The lattice constants of one, two, and three atomic layers of TiN are taken to be the same as the bulk instead of their stress-free values (see the Supplementary Materials for explanations). In addition to the stand-alone monolayer mentioned above, we construct two additional monolayer structures: a monolayer of TiN sandwiched in bulk by Ar above and below, and a surface monolayer separated from bulk by a layer of Ar. In both cases, the Ar atoms are aligned with the hollow site both of the monolayer and of the slab surface layer. The periodic superlattice structure contains five layers of TiN alternating with the Ar-TiN-Ar sandwich. The Ar-TiN top layer structure consists of a five-layer TiN slab, the Ar-TiN top layer, and above that a vacuum of at least 20Å.

In this study, we perform first-principles calculations within the framework of density functional theory (DFT), in the all-electron full-potential-linearized augmented-plane-wave (FLAPW) basis (*35*), with the exchange-correlation potential taken under the Perdew-Burke-Ernzerhof (PBE) generalized gradient approximation (*36*). The interlayer spacings of the bilayer and trilayer are further relaxed until all the forces between atoms reduce below 2mRy/bohr. For the Ar-inserted structures, the structure relaxations are instead performed in the projector-augmented wave (PAW) basis with van der Waals force included (*37-40*). An additional onsite Coulomb repulsion $U$ is added to the Ti 3$d$ orbital (to be called PBE+$U$ throughout this paper) in the FLAPW basis for calculating electronic structures, where the $U$ values are calculated using the constraint-PBE method (*32*). The electronic calculations of all above four TiN structures are used to further calculate their valence charges and spins by Bader analysis.

We also calculate the exchange couplings between Ti atoms of the TiN monolayer. The Heisenberg Hamiltonian ($H = J S_1 \cdot S_2$) provides the simplest estimation of those couplings, where the values of $J$ is extracted out of the energy differences between the ferromagnetic (FM) and antiferromagnetic (AFM) systems. We need to consider two antiferromagnetic phases in the FM-AFM energy difference in order to calculate the dominating $J$'s that couple two nearest-neighbored and next-nearest-neighbored Ti atoms, namely $J_1$ and $J_2$ respectively (see Fig. 2). After carefully counting the couplings that occur in the unit cell of each antiferromagnetic phase, we relate $J_1$ and $J_2$ to the DFT-calculated transition energies between FM and AFM as follows:

$$\Delta E_1 = E_{FM} - E_{AFM1} = 8 J_1 \times |S|^2 \quad (1)$$

$$\Delta E_2 = E_{FM} - E_{AFM2} = 4 J_1 \times |S|^2 + 8 J_2 \times |S|^2 \quad (2)$$

Here $S$ is the magnitude of spin, $E_{FM}$ and $E_{AFM}$ are the total energies of FM and AFM phases calculated by DFT, respectively.

## H2: Supplementary Materials

Strained monolayer and bilayer
Magnetism of Ti in different surroundings
The effect of spin-orbit coupling in a monolayer
fig. S1. Structures of Ar-inserted TiN.
fig. S2. Ti 3$d$ PDOS of TiN monolayer in different structures.
fig. S3. Model structures adapted from a TiN bilayer.
fig. S4. Bandstructures and Ti 3$d$ partial densities of states of a TiN monolayer, with and without spin-orbit coupling in either x or z direction.
Table S1. The calculated magnetic moments in three different structures as in fig. S2.
References (*41, 42*)

**Acknowledgments**:
We thank R. Pushpa and S. Gangopadhyay for their stimulating discussions. **Funding:** C.-Y.L. and S.-W.Y. are partially funded by Taiwan Ministry of Science and Technology grants MOST 104-2112-M-009-005, MOST 104-2119-M-009-008 and MOST 105-2119-M-009-009, and acknowledge the facility support from the Taiwan National Center for High-Performance Computing. K.-L.O. and S.-W.Y. acknowledge financial support from the Taiwan Ministry of Health and Welfare grants MOHW103-TDU-N-211-133001. The work of B.A.J. was performed in part at the Aspen Center for Physics, which is supported by National Science Foundation grant PHY-1066293. **Author contributions:** C.-Y.L. instructed the details of all scientific computing tasks and has done the majority of the manuscript preparation. S.-W.Y. prepared and maintained these computing tasks and collected all computed data. K.-L.O. proposed a future biomedical application of this study and directed the studies of electrical properties in addition to magnetism. B.A.J. provided physical interpretations of both the monolayer magnetic phases and the quenching of magnetism of the bilayer and provided the originating idea for the study.
**Competing interests:** The authors declare that they have no competing interests.
**Data and materials availability:** All data needed to evaluate the conclusions in the paper are present in the paper and/or the Supplementary Materials. Additional data related to this paper may be requested from the authors.

**Figures and Tables**

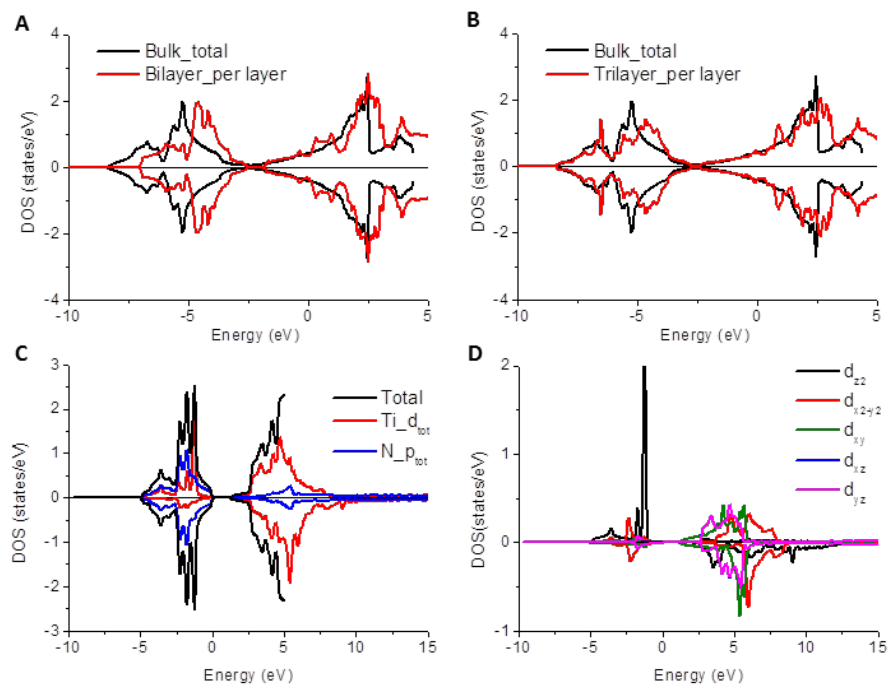

**Fig. 1. The densities of states of TiN structures.** (**A**) The total DOS of TiN bulk and the DOS per layer of the bilayer structures. (**B**) The total DOS of TiN bulk and the DOS averaged per layer of the trilayer structures. (**C**) For the monolayer, the total density of states of the *d*-electrons of Ti, the *p*-electrons of N, and the total electrons in the unit cell, including interstitial contributions. (Values of the total above 5eV have been removed because the interstitial contributions are large and are not part of the current discussion.) (**D**) The Ti 3*d* PDOS of TiN monolayer structure. For all four figures, positive (negative) refers to spin-up (spin-down).

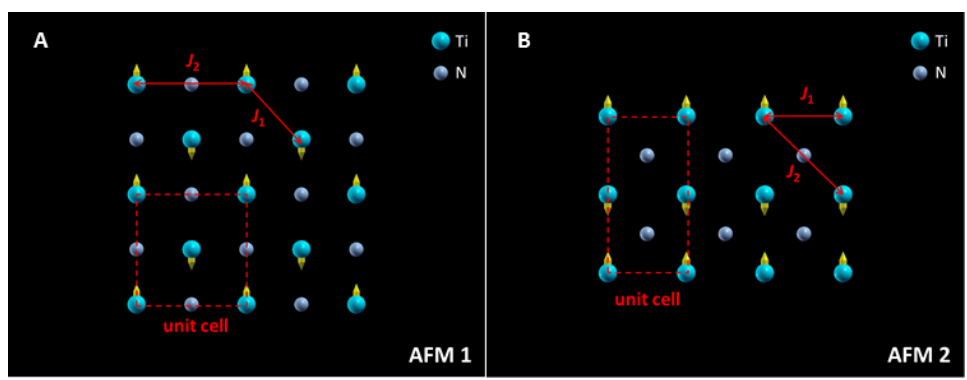

**Fig. 2. Schematic diagrams showing antiferromagnetic spin configurations in a TiN monolayer.** (**A**) The antiferromagnetic phase 1 (AFM1): the opposite spins are associated with two square sublattices, respectively. (**B**) The antiferromagnetic phase 2 (AFM2): the opposite spins are associated with alternating stripes. For both figures, the red dashed border lines specify the unit cells, and the yellow arrows indicate the spin directions. The exchange couplings $J_1$ and $J_2$ are associated with the Ti atoms as indicated.

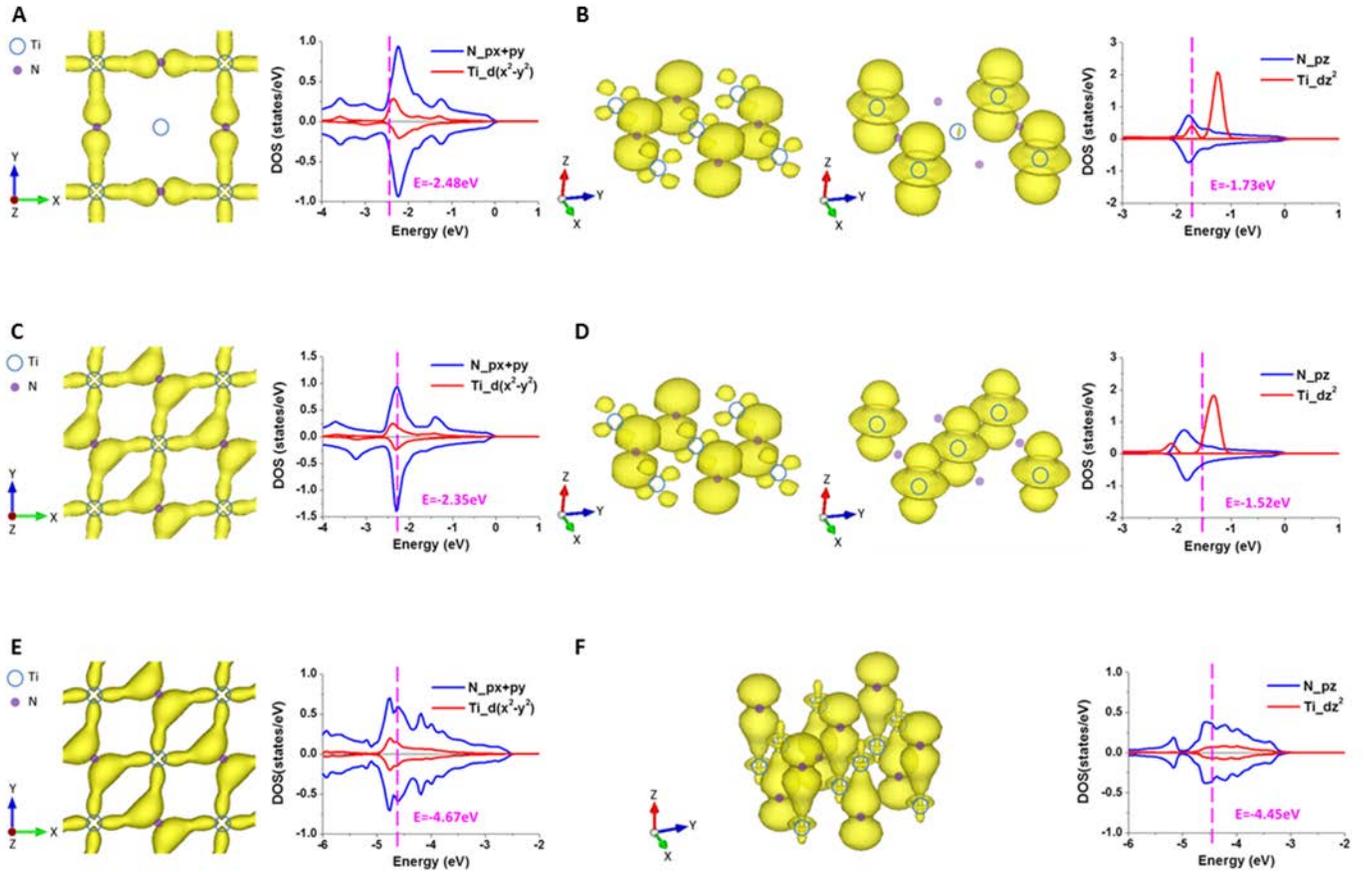

**Fig. 3. Isosurface plots of the Ti-N binding orbitals for TiN monolayer and bilayer.** Three plotted systems include two of the TiN-monolayer magnetic phases, AFM1 in (A and B) and FM in (C and D), and the nonmagnetic bilayer in (E and F). The vertical purple dashed lines in all PDOS plots indicate the energy of their associated orbitals. (**A**) One representative inplane binding orbital; the PDOS shows it has a N $p_x, p_y$ + Ti $d_{x^2-y^2}$ binding nature. (**B**) Two representative out-of-plane dangling orbitals; the PDOS shows the two degenerate orbitals have orbital natures of N $p_z$ + Ti ($d_{xz}+d_{yz}$) and Ti $d_{z^2}$, respectively. (**C**) One representative inplane binding orbital; the PDOS shows a N ($p_x + p_y$) + Ti $d_{x^2-y^2}$ binding nature. (**D**) Two representative out-of-plane, degenerate dangling orbitals; the PDOS shows they have orbital natures of N $p_z$ + Ti ($d_{xz}+d_{yz}$) and Ti $d_{z^2}$, respectively. (**E**) One representative inplane binding orbital; the PDOS shows a N ($p_x + p_y$) + Ti $d_{x^2-y^2}$ binding nature. (**F**) One representative out-of-plane binding orbital; the PDOS shows a N $p_z$ + Ti $d_{z^2}$ binding nature.

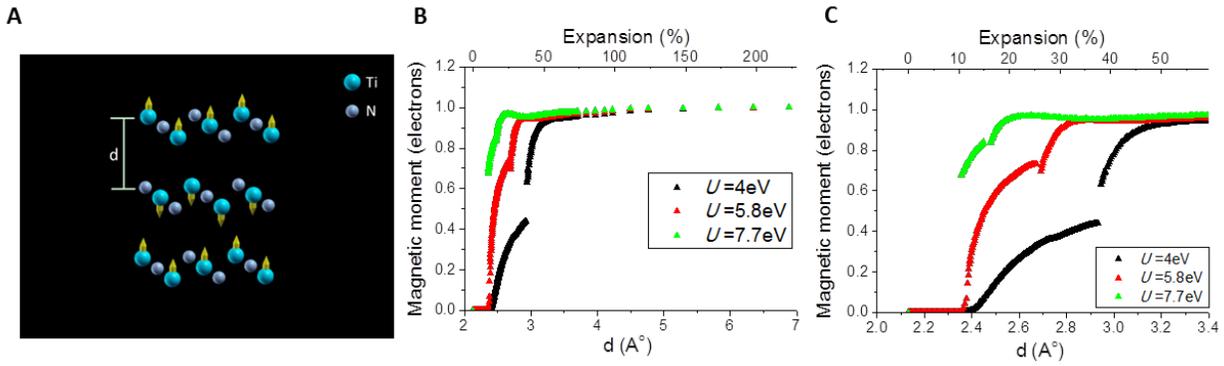

**Fig. 4. Predicted magnetic transition due to a uniaxial expansion on TiN bulk.** (**A**) Schematic plot of a uniaxially-expanded TiN bulk, where *d* is the layer spacing. The yellow arrows indicate the direction of the Ti spins. (**B** and **C**) The calculated magnetic moments per unit cell per layer, as functions of *d* with *U*=4, 5.8, 7.7eV, respectively. The upper axis shows the respective percentage of expansion in *d*. In (**C**), we have a closer look at the transition in (B).

| TiN Structure | $U$ (eV) | Magnetic moment | |
|---|---|---|---|
| | | Ti | N |
| Bulk | 4.0 | 0.00 | 0.00 |
| Monolayer | 7.7 | 1.11 | 0.00 |
| Bilayer | 7.2 | 0.00 | 0.00 |
| Three layers | 7.2 | 0.00 | 0.00 |

**Table 1. The calculated magnetic moments of TiN bulk, monolayer, and bilayer.** The unit of magnetic moment is number of electrons. $U$ is the coulomb repulsion applied to the Ti $3d$ orbital.

# Supplementary Materials

**Strained monolayer and bilayer**

The equilibrium lattice constants of the TiN monolayer and bilayer can be calculated by DFT using the PBE exchange correlation functional. Calculations show these two-dimensional lattice constants are 5.6% and 3.2% smaller than the bulk one, respectively. However our present computational studies do not intend to suggest experiments in a few free-standing layers of TiN, which would be difficult to construct, but instead on top of a substrate. Therefore, we model such atomic layers under a uniform inplane strain to maintain a lattice constant same as the bulk. Experimentally, such a condition can even be precisely controlled by alternatingly depositing molecules and TiN on a substrate, or by separating layers molecularly. In order to illustrate such experimental constructions, we perform additional DFT calculations of two model systems with Ar atoms inserted into the TiN interstitials; one has a TiN monolayer sandwiched between Ar within the bulk, and the other has Ar below the surface monolayer (see fig. S1). The comparisons of Ti PDOS and magnetic moment in the monolayer formed with/without Ar inserted are shown in fig. S2 and Table S1, respectively. One can see that a TiN monolayer has essentially the same magnetism whether it is next to a layer of Ar.

**Magnetism of Ti in different surroundings**

In the main part of this paper, we find that the magnetism of a monolayer of TiN disappears as another layer of TiN approaches it. In this supplement we study a variety of subsurfaces for a layer of TiN, and show the presence of magnetism to be strongly dependent on whether extensive hybridization can take place to distort the $d$-orbitals. This can cause the extra $d$ electron, originally giving rise to the magnetism, to be subsumed into the interlayer region, mixing with other $d$ and $sp$ orbitals.

We first note that the magnetism of a single layer of TiN is not affected by N coupling via $xy$ bonding in the same layer, as the Ti orbitals mostly responsible for magnetism are in the $z$-like directions. This understanding can be extended to a related system. Pushpa et al studied a layer of TiN on a Cu substrate (*41*). In this system the Ti hybridization with the N is in-plane, and no magnetism hybridizes to the interlayer region. The N themselves are weakly magnetized. In this case, the sublayer Cu acts mainly as a locally spin-polarized sea of electrons. For the Ti, only in the extreme case of a very low-temperature Kondo effect could these electrons screen the spin. For the N, the Cu electrons provide another direction of coupling and satisfy many of the open bonds on the N. Pushp also studied a model system in which the Ti was moved to directly over the N, which causes the N to sink down into the copper surface. In this case, the vertical Ti-N coupling enhanced the Ti spin over its value when in a TiN layer, an effect we will also see below, in our second model system.

When a second layer or subsurface is added with open shell atoms, such as N, the $z$-orbitals get occupied, which can give rise to completely different bonding on the Ti. The interplay between magnetism and bonding, in which too strong covalent bonding can cause the loss of magnetism, is a characteristic of our TiN bilayer, in which the strong vertical coupling between the Ti and the N below it causes the loss of magnetism on the Ti. The competition between magnetism and bonding appears in another system, Ni adatoms on MgO, in reference (*42*). There, when bonding became too strong (covalent Ni-O for Ni at the O site), the Ni loses its magnetism. When the Ni is on the Mg (weak bonding), the Ni magnetism is restored.

We further study this competition between magnetism and bonding with some model bilayer systems that illuminate the magnetism. In the first, we remove the Ti from the layer below, leaving just the N, so that every Ti is above a N, and every N in the TiN layer has no other atom directly below it (see fig. S3). In this case, the same bonding that occurs with TiN bilayers continues to occur, and the Ti lose their magnetism as the layer of N approaches closer than a certain distance. However, in addition, interestingly, because the unoccupied orbitals of the bare N are no longer *xy* bonding with the Ti, they develop large moments of their own. (N has a half-filled p-shell, and so in the right conditions can develop spins as large as 3/2.)

The second model system removes the N from the second layer, leaving a monolayer of elemental Ti. In this system, the bare Ti are arranged so that each is opposite an N in the layer below it. The Ti in the TiN layer have no atom directly above them (see fig. S3). Now the N in the TiN layer are bonding in both *xy* and *z* directions, which changes their bonding with Ti in the same layer. The N-Ti vertical combination, at a moderate distance, results in a large moment on the Ti, just as in the case when Ti is an adatom on the CuN surface, described above. The altered bonding on the N results in a net *z*-component to its bonding to the surface Ti, and reduces the Ti's moment.

To show that the N atoms, and their bonding, are the active players in reducing (or enhancing) the magnetism of the Ti, we finally point to a system in which all N have been removed, bulk metallic Ti, which is paramagnetic, with each Ti maintaining a magnetic moment even with layers above and below.

**The effect of spin-orbit coupling in a monolayer**

We include the spin-orbit coupling in calculating a monolayer, and plot the band structures and Ti 3*d* PDOS in fig. S4 to compare them with those without the spin-orbit coupling.

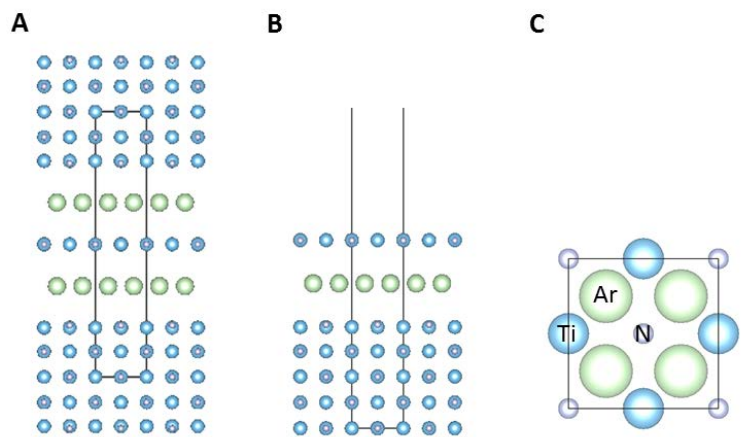

**fig. S1. Structures of Ar-inserted TiN.** (**A**) A monolayer of TiN sandwiched in bulk by Ar above and below. The distances from Ar to monolayer and to bulk are 3.471 and 3.448Å, respectively. (**B**) A surface monolayer separated from bulk by a layer of Ar. The distances from Ar to monolayer and to bulk are 3.563 and 3.591Å, respectively. (**C**) Top view of both (**A**) and (**B**) where only the monolayer of TiN and Ar are displayed. In all three figures, the black border lines specify the unit cells.

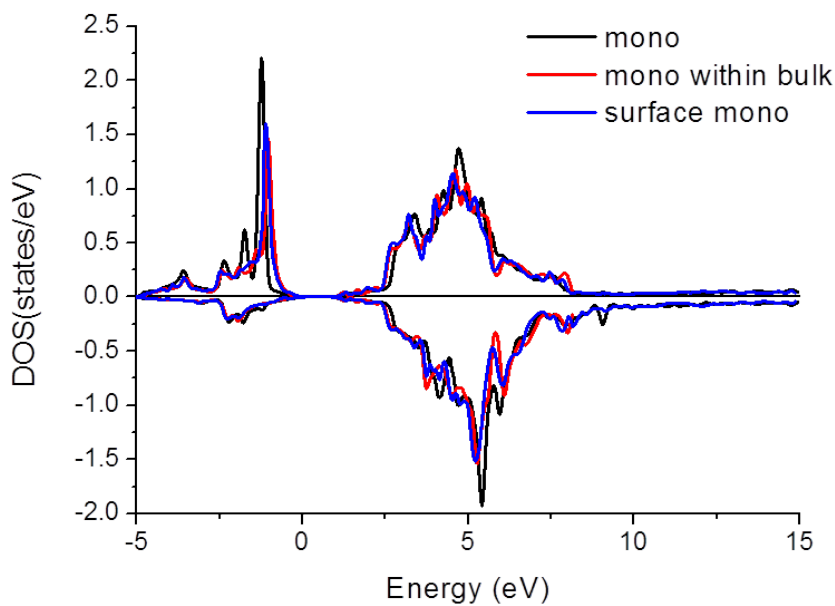

**fig. S2. Ti 3*d* PDOS of TiN monolayer in different structures.** The labels of the curves "mono", "mono within bulk", and "surface mono" stand for the stand-alone monolayer, the monolayer as in fig. S1A, and the one as in fig. S1B, respectively. For all curves, positive (negative) refers to spin-up (spin-down).

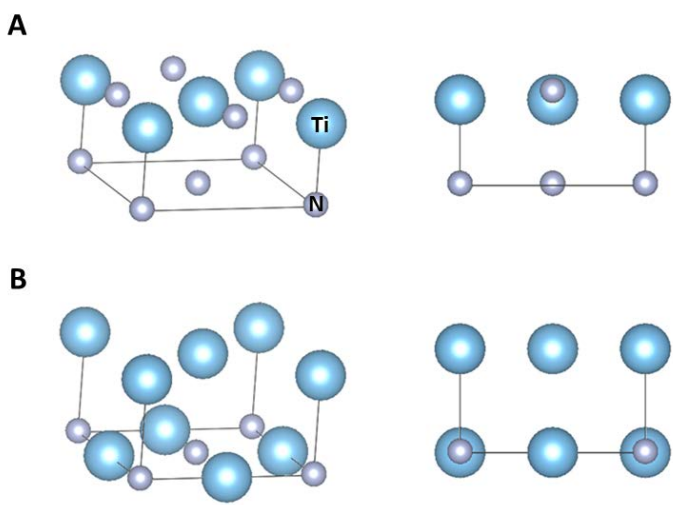

**fig. S3. Model structures adapted from a TiN bilayer,** (**A**) by removing Ti atoms from one particular atomic layer, and (**B**) by removing N atoms from one layer. In both (**A**) and (**B**), the oblique views on the left and side views on the right. The square at the bottom of each left figure denotes the unit cell along the parallel, and the vertical line segments indicate the vertical bonds.

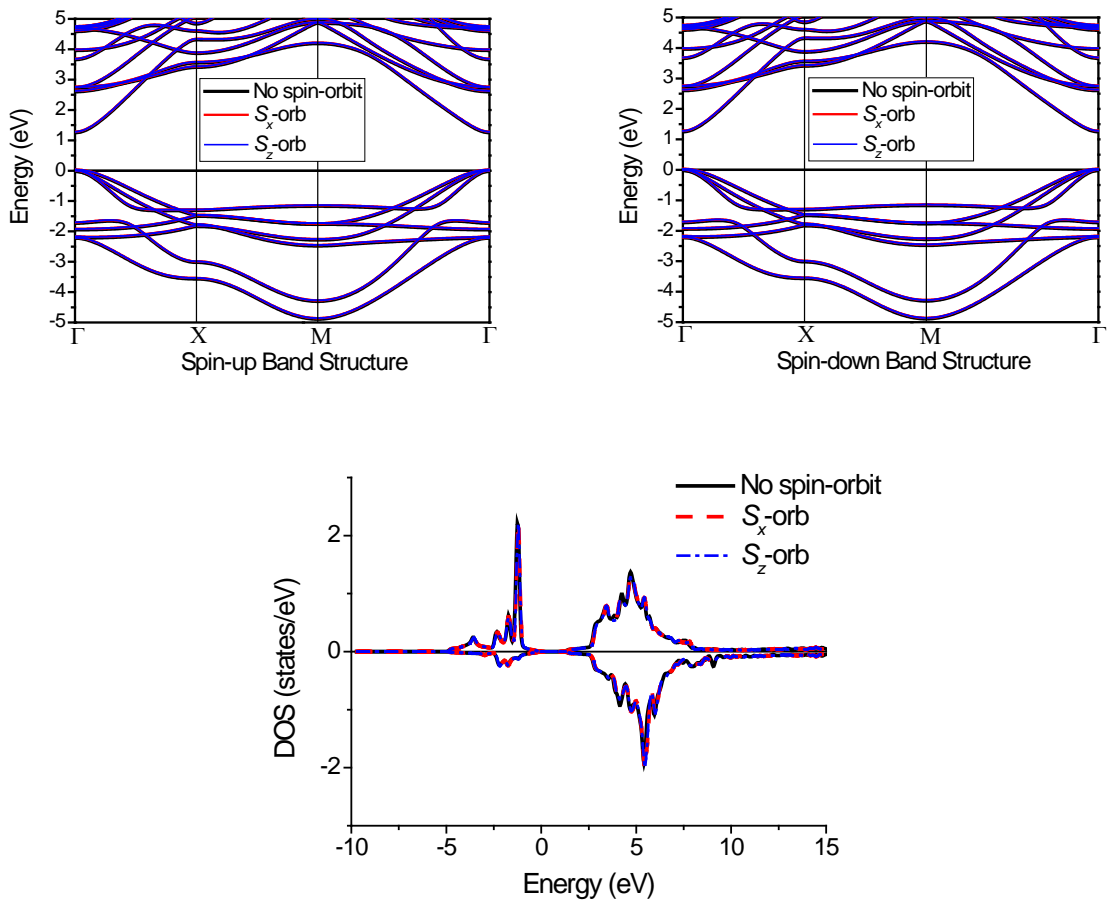

**fig. S4. Bandstructures (upper) and Ti 3d partial densities of states (lower) of a TiN monolayer, with and without spin-orbit coupling in either *x* or *z* direction.** Note that the effect of the spin-orbit coupling is negligible on both figures.

|  | Ti Magnetic moment |
|---|---|
| Monolayer | 1.11 |
| Mono within bulk | 1.10 |
| Surface mono | 0.93 |

**Table S1. The calculated magnetic moments in three different structures as in fig. S2.** The unit of magnetic moment is number of electrons. The coulomb repulsion strengths applied to the Ti 3*d* orbital are $U = 7.7$ eV in all three cases.